\documentclass[useAMS,usenatbib]{mn2e}
\usepackage{psfig}
\newcommand{\um}{\hbox{$\mu {\rm m}$}}
\newcommand{\uJy}{{\rm\thinspace \mu Jy}}
\newcommand{\fluxerg}{\hbox{$\erg\cm^{-2}\s^{-1}\,$}}

\newcommand{\erg}{{\rm\thinspace erg}}
\newcommand{\km}{{\rm\thinspace km}}
\newcommand{\cm}{{\rm\thinspace cm}}
\newcommand{\s}{{\rm\thinspace s}}
\newcommand{\kmpspMpc}{\hbox{$\km\s^{-1}\Mpc^{-1}\,$}}
\newcommand{\Mpc}{{\rm\thinspace Mpc}}
\newcommand{\mJy}{{\rm\thinspace mJy}}
\newcommand{\eg}{{e.g.}}
\newcommand{\ie}{{i.e.}}
\newcommand{\bfig}{\begin{figure}}
\newcommand{\efig}{\end{figure}}

\newcommand{\aap}{A\&A}

\newcommand{\aj}{AJ}
\newcommand{\apj}{ApJ}
\newcommand{\apjl}{ApJL}
\newcommand{\apjs}{ApJS}

\newcommand{\araa}{ARA\&A}
\newcommand{\mnras}{MNRAS}

\title[The Star Formation History of the Universe as Revealed by Deep Radio Observations]{The Star Formation History of the Universe as Revealed by Deep Radio Observations}
\author[N. Seymour et al.]{N. Seymour$^{1}\thanks{E-mail: seymour@ipac.caltech.edu}$, T. Dwelly$^{2}$, D. Moss$^{2}$, I. M$^{\rm c}$Hardy$^{2}$, A. Zoghbi$^{2,3}$, G. Rieke$^{4}$, \and M. Page$^{5}$, A. Hopkins$^{6}$ and N. Loaring$^{7}$\\
$^{1}${\it Spitzer} Science Center, Caltech, 1200 East California Boulevard, Pasadena, CA 91125, USA. \\
$^{2}$School of Astronomy \& Astrophysics, University of Southampton, Highfield, Southampton, SO17 1BJ, UK. \\
$^{3}$Institute of Astronomy, University of Cambridge, Madingley Road, Cambridge, CB3 0HA, UK.\\
$^{4}$Steward Observatory, Tucson, USA.\\
$^{5}$Mullard Space Science Laboratory, UCL, Holmbury St. Mary, Dorking, Surrey, RH5 6NT, UK. \\
$^{6}$University of Sydney, Australia. \\
$^{7}$SALT, PO box 9, Obsevatory, 7925, South Africa.}
\begin{document}

\date{Draft $19^{\rm th}$ February 2008}


\maketitle

\label{firstpage}

\begin{abstract}
Discerning the exact nature of the sub-mJy radio population has been 
historically difficult due to the low luminosity of these sources at most 
wavelengths. Using deep ground based optical follow-up and observations from 
the {\it Spitzer Space Telescope} we are able to disentangle the 
{\em radio-selected} Active 
Galactic Nuclei (AGN) and Star Forming Galaxy (SFG) populations for the first 
time in a deep {\em multi-frequency} VLA/MERLIN Survey of the $13^{\rm H}$ 
{\it XMM-Newton/Chandra} Deep Field. The discrimination diagnostics include 
radio morphology,  radio spectral index, radio/near-IR and mid-IR/radio flux 
density ratios. We are now able to calculate the extragalactic 
Euclidean normalised source counts separately for AGN and SFGs. We find 
that while SFGs dominate at the faintest flux densities and 
account for the majority of the up-turn in the counts, AGN still make up 
around one quarter of the counts at $\sim50\,\uJy$ (1.4\,GHz). Using 
radio luminosity as an unobscured star formation rate (SFR) measure we are 
then able to examine the comoving SFR density of the Universe up to $z=3$ 
which agrees well with measures at other wavelengths. 
We find a rough correlation of SFR with stellar mass for both the sample 
presented here and a sample of local radio-selected SFGs from the 6df-NVSS 
survey. This work also confirms 
the existence of, and provides alternative evidence for, the evolution of 
distribution of star formation by galaxy mass: ``downsizing''. As both these
samples are SFR-selected, this result suggests that there is a maximum SFR 
for a given galaxy that depends linearly on its stellar mass.
The low ``characteristic times'' (inverse specific SFR) of the SFGs in 
our sample are similar to those of the 6dF-NVSS sample, implying that most
of these sources are in a current phase of enhanced star formation.
\end{abstract}


\section{Introduction}

Whilst observations of powerful, radio-loud Active Galactic Nuclei (AGN) were 
an early probe of the distant Universe \cite[see][for a summary]{Stern:99e}, 
starburst galaxies are three or more orders of magnitudes less luminous at 
radio wavelengths and hence difficult to observe at large distances. However 
the deepest radio surveys at 1.4\,GHz now reach an rms below $10\uJy$ 
\citep[][etc.]{Richards:00,Seymour:04, Biggs:06, Fomalont:06}. These deep 
radio observations reveal a (very well characterised) up-turn in the Euclidean 
normalised sources counts below 1\,mJy above that predicted from the 
extrapolation of the AGN counts measured at brighter flux densities. This 
up-turn has been attributed to the emergence of a star forming galaxy (SFG) 
population, requiring strong evolution of the SFG radio luminosity function 
\citep{RowanRobinson:93,Hopkins:98, Seymour:04, Moss:07}, 
although some authors argue that there is a significant contribution due to 
relatively weak radio AGN \citep{Simpson:06, Huynh:07a, Barger:07}. 

Determining the nature of individual radio sources has remained difficult due 
to their low luminosities at other wavelengths. Discrimination between 
radio-selected SFGs and AGN in these deep surveys would, 
for example, allow an {\em independent} measure of the star formation history 
of the Universe as star formation rate (SFR) is directly related to radio 
luminosity for galaxies with weak or no AGN radio emission \citep{Condon:92}. 
The comoving SFR density (SFRD)
has been determined previously from deep radio data out to $z\sim 1.6$
\citep{Haarsma:00}, but the study presented here offers  
several improvements and advantages, noticeably a more robust and systematic 
discrimination between AGN and SFG, the use of more accurate redshifts,
an extension to higher redshift and a larger area/sample size by 
a factor of $\sim5$.

Whilst radio surveys have not traditionally been used to determine SFRs in the
distant Universe, other methods have their limitations. The far-IR and sub-mm 
wavelength range is limited by the relatively poor sensitivity and angular 
resolution of current instrumentation and telescopes. The mid-IR, whilst 
very sensitive, is not as reliable a tracer of SFR at high redshift as 
the far-IR and sub-mm \citep[e.g.][]{Papovich:07}. UV emission from young 
massive stars can easily be obscured by dust, but the extrapolation of models 
derived from local galaxies likely break down for the more luminous galaxies,
$>L_*$, found in deep optical surveys. Furthermore, the rest-frame 
UV is redshifted across a large wavelength range (UV/optical/near-IR) at 
cosmological distances so different detectors are needed to trace the UV at 
different redshifts. Determination of the SFR from optical emission lines 
is also affected by obscuration and shifting to longer wavelengths and hence 
in and out of observable wavelength windows. The optical emission lines can 
also suffer from absorption by dust. X-ray emission associated with star 
formation (e.g. from binary stars) is intrinsically weak and is subject to 
varying degrees of photoelectric absorption. Even at 
the faintest X-ray fluxes reached in the CDFN ($\sim10^{-17}\,\fluxerg$), 
the source counts are dominated by AGN \citep{Bauer:04}. Given that 
forthcoming radio facilities (e.g EVLA, e-MERLIN, LOFAR, ASKAP, and 
eventually the SKA) will reach many orders of 
magnitude deeper than the current deepest radio survey, and will be able to 
detect star forming galaxies out to the era of re-ionization ($z\la7$), 
it is timely to consider how to characterise the emission from 
faint radio sources and which supporting data are the most valuable.

We make an important distinction here between radio indicators of an AGN and 
{\em non-radio} indicators of an AGN. We need to separate the radio sources 
into those whose radio emission is AGN dominated and those that are consistent 
with being dominated by star formation. By AGN powered indicators we mean some 
measure that indicates that 
the radio emission from a galaxy is dominated by accretion onto a 
super-massive black hole and/or from associated jets and lobes (\ie\ 
radio-loud AGN). We assume, for simplicity, that the dominant power in most 
sources is either from an AGN {\em or} from star formation, but otherwise 
make no a priori assumptions about the sources, \eg\ the SED.
The radio AGN indicators we use in this paper are radio/near-IR flux density 
ratio, mid-IR/radio flux density ratio, radio morphology and radio spectral 
index. We do not include {\em non-radio} AGN indicators, \eg~classical methods 
like optical/IR emission lines, optical morphology, X-ray observations etc. 
and methods from recent results with the {\it Spitzer Space Telescope} based on 
mid-IR SEDs \citep[e.g. ][]{Donley:05,Stern:05b,Lacy:05}. As these non-radio 
indicators tell us 
nothing directly about the nature of the radio emission we do not use them in 
our discrimination methods. A comparable philosophy was previously adopted by 
\citet{Muxlow:05}.

In this paper we will present our method for classifying the faint radio
sources and examine the properties of the SFG population.
The radio, optical and IR observations are summarised in section 2. The 
various methods of discriminating between AGN and SFGs are presented in 
section 3. In section 4 we present the Euclidean 
normalised source counts separately for AGN and SFGs for the first time. 
In section 5 we study the general properties of the radio-selected SFG 
population, present the comoving SFR density of the Universe up to 
redshift $\sim3$ derived from our data, examine
the distribution of star formation with host galaxy stellar mass and derive 
characteristic times. We conclude this article in section 6.
Throughout we use a concordance model of Universe expansion, 
$\Omega_M = 1 - \Omega_{\Lambda} = 0.3$, $\Omega_0 = 1$, and 
$H_0 = 70\, \kmpspMpc$ \citep{Spergel:03}. Magnitudes are AB unless otherwise 
stated.

\section{The data}

The 13$^{\rm H}$ {\it XMM-Newton/Chandra} Deep Survey field presents us with a 
unique data set for performing a radio-based SFG/AGN separation of radio 
sources. This field,
centred at 13$^{\rm h}\,34^{\rm m}\,37^{\rm s}$ +37$^\circ\,54'\,44''$, was 
the location of one of the deepest {\it ROSAT} surveys \citep{McHardy:98b}, 
and lies in a region of 
extremely low Galactic absorption ($N_H\sim 6\times10^{19}\,$cm$^{-2}$). 

\subsection{The radio data}

\subsubsection{The VLA 1.4\,GHz observations}

The $13^{\rm H}$ {\it XMM-Newton/Chandra} deep field was observed by the VLA 
at 1.4\,GHz for 14\,hrs in A-array in August 1996 and 10\,hrs in B-array in 
November 1995. 
A catalogue of 449 sources above $30\,\uJy$ ($4\,\sigma$ detection limit) in 
a $30$\,arcmin diameter circular field of view (0.196\,deg$^2$) was obtained 
with a resolution of $3.3$\,arcsec.
The Euclidean normalised source counts, corrected for the 
incompleteness of the catalogue due to instrumental effects, were presented 
in \citet[][hereafter S04]{Seymour:04}. The 449 sources detected at $1.4\,$GHz 
with the VLA constitute the parent sample on which this work is based. In 
contrast to S04 we use the peak flux density instead of total flux 
density for sources that are significantly resolved and at low S/N as the peak
flux is a better measure of the true flux in these cases. 

The size of this sample (in terms of area and number of sources) is well 
positioned between those fields of the similar area  but are slightly deeper  
\citep[by $\sim40\%$ \eg\ the Lockman Hole and the HDFN,][]{Biggs:06} and 
those of larger area \citep[\eg~AEGIS and COSMOS,][]{Ivison:07a,Schinnerer:07}, 
but which are less deep by $\sim40\%$. 

\subsubsection{The MERLIN 1.4\,GHz observations}

The $13^{\rm H}$ field was observed with Multi-Element Radio-LInked Network in 
April 1999. Due to the smaller field of view, $10$\,arcmin, four MERLIN 
pointings were used to cover most of the VLA field of view. 
MERLIN has a higher resolution, $0.2$\,arcsec, than the VLA 
due to its greater maximum base-line. We used the parent catalogue as a 
detection list and only imaged the positions around known sources (including
bright sources not in the principle VLA or MERLIN field of view). We made 
MERLIN images for all VLA detections with $S_{\rm 1.4GHz}>72\,\mu$Jy that lay 
inside the MERLIN coverage. These images were combined by first making 
maps without the beam deconvolved and beams around each source in the 
sky plane, then averaging the maps and beams  before deconvolving the 
average beam from the average map as described by A. Zoghbi et al. 
(in prep.) The combination of datasets from different telescope arrays is 
non-trivial as the four MERLIN pointings are offset from the VLA phase center
and both the VLA and MERLIN images suffer from radial smearing. However
for sources in the parent sample having flux densities above $\sim100\,\uJy$, 
and that are favourably positioned, we can make high-resolution maps from 
combined MERLIN and VLA data.

\subsubsection{The VLA 4.8\,GHz observations}

The $13^{\rm H}$ field was observed with the VLA at 4.8\,GHz in April 1991. 
The observations cover the whole field of view of the 1.4\,GHz data, 
$30$\,arcmin \citep[observations originally designed to match the {\it ROSAT} 
field of view][]{McHardy:98b} with 51 separate pointings. The observations 
reach $33\,\uJy$ rms, and were performed in VLA D-array hence the restoring 
beam is $14$\,arcsec, around a factor four greater than the $1.4\,$GHz 
observations. From the catalogue of \citet{Seymour:02} we find 45 sources 
with flux densities 
$>4\,\sigma$ and within the 30\,arcmin diameter field of view of the 1.4\,GHz 
observations. Of these sources 45 have counterparts in the 1.4\,GHz 
catalogue within $7$\,arcsec and hence we can calculate their $1.4-4.8\,$GHz
spectral indices. We define the radio spectral index, $\alpha$, by 
$S_\nu\propto\nu^\alpha$, and hence $\alpha=-1.85\log_{10}(S_{1.4}/S_{4.86})$.
We use total, rather than peak, flux densities when calculating the radio 
spectral index. In our earlier study (S04), we made tapered maps of the 
1.4\,GHz data but we did not find significant extended radio emission that 
could affect the calculated spectral 
indices. There is one instance where two nearby 1.4\,GHz sources (separated by 
$20$\,arcsec) are matched to a single 4.8\,GHz source and therefore we cannot 
calculate the spectral indexes separately for these two sources.

\subsection{Optical, near-IR and mid-IR data}
\label{sec.opdat}

\subsubsection{Optical and near-IR data}

We have obtained imaging in many bands from the near-UV to near-IR
over recent years: $u^*$, $g'$ and $i'$-band from
MegaCam/CFHT, $B$, $R$, $I$ and $z'$-band from SuprimeCam/Subaru, $Z$-band from
WFC/INT, $J$-band from WIRC/Palomar, $H$-band from WIRCam/CFHT, and $K$-band
from WFCAM/UKIRT. This imaging covers the entire parent sample bar the
WIRC $J$-band (which covers approximately 3/4 of the survey area), and
has been supplemented by hundreds of optical spectra of the
counterparts to our radio and X-ray sources. Of our parent sample
164/449 currently have redshifts determined from optical
spectroscopy. 

\subsubsection{Mid-IR data}

The $13^{\rm H}$ field was observed by the IRAC \citep{Fazio:04a} and MIPS 
instruments \citep{Rieke:04} on board the {\it Spitzer Space Telescope} 
\citep{Werner:04} in July 2005 as part of MIPS instrument team GTO time 
(PI G. Rieke, program identification number 81). A strip of 
$\sim0.5\times 1\deg$ containing the $13^{\rm H}$ field was imaged by both 
instruments. The IRAC data consisted of observations with all four channels 
($3.6, 4.5, 5.6$ and $8.0\,\um$) reaching $1\,\sigma$ flux density limits of 
0.7, 1.2, 7.2 and $7.6\,\uJy$ respectively. 
These values represent the limiting flux  density of
a point source for which we can make a flux measurement in a $3.8$\,arcsec
diameter aperture with $>1\sigma$ accuracy.
Most, 388/449 ($87\%$), of our 
parent sample were detected in at least the $3.6\,\um$ channel.
The $24\,\um$ data was taken as part of a simultaneous scan map in each of the 
three bands (24, 70 and $160\,\um$) 
reaching a $3\sigma$ detection limit of $98\,\uJy$ in a 10.5\,arcsec aperture. 
We found 330/449 ($73\%$) of our parent sample had
$24\,\um$ counterparts. We do not use the 70 and 160\,$\um$ data
here as they are not sensitive enough to provide information on 
most of the radio sources, and then only those at low redshift.

\subsubsection{Photometric Redshifts}

We have used our multi-band near-UV to near-IR together with IRAC
3.6, 4.5 and $5.8\,\um$ data to calculate photometric redshifts for the
optical counterparts to our radio sources.  Full details of this
process will be presented in Dwelly et al. (in prep. hereafter D08), 
but here we give a short summary.
Initial catalogues were generated independently for each optical,
near-IR and mid-IR waveband using Sextractor \citep{Bertin:96}.
These initial catalogues are then combined into
a single ``master'' catalogue containing only unique detections. 
Aperture photometry is carried out in each waveband at the
locations of each master catalogue source. In the optical/near-IR bands
we used a 3\,arcsec diameter aperture for bright sources ($i'<20$), and
 a 2\,arcsec aperture for fainter sources to improve
S/N. In the IRAC bands a 5.8\,arcsec diameter aperture was employed.
Appropriate aperture corrections are applied given the different point
spread functions in each waveband \citep{Gawiser:06}.
We match optical/near-IR counterparts to the parent radio sample are by 
searching for all objects in the ``master'' catalogue lying within 1.5\,arcsec 
of the peak of the radio emission. Where available we use the high resolution 
images from combined MERLIN+VLA maps to determine the correct optical 
counterparts. We examine the optical and radio images for sources with more 
than one counterpart, and manually choose the most appropriate object.  
We then used the publicly available {\tt Hyperz} code \citep{Bolzonella:00} 
to determine photometric redshifts for each source from their multi-band 
aperture photometry. Only the wavebands shortward of $5.8\,\um$ were used to 
determine photometric redshifts in order to avoid complications with PAH 
features. We used the standard set of synthetic evolving galaxy templates 
supplied with {\tt Hyperz}. To ensure reliability, we ignore the photometric 
redshifts calculated for all radio sources detected with $>3\,\sigma$ 
significance in fewer than 4 wavebands. Hence, 60 radio sources from our 
parent sample have no photometric redshift information including 31 with no
counterparts detected in any optical, near-IR or mid-IR waveband. Most of the 
other 29 sources are detected only in the $K-$band data and/or the two shorter 
wavelength IRAC channels, and hence the derived photometric redshifts have 
very large uncertainties. We
consider the likely nature of these objects in section 3.6.1.

\subsection{Local 6dF-NVSS comparison sample}

To illustrate our discrimination techniques we use the 6dF-NVSS sample of local
spectroscopically identified radio sources from \citet{Mauch:07}. 
These authors identified $\sim8000$ radio sources from the 1.4\,GHz NRAO VLA 
Sky Survey \citep[NVSS,][]{Condon:92} in the Second Incremental Data Release 
of the 6 degree Field Galaxy Survey (6dFGSDR2) with galaxies brighter than 
$K = 12.75$ (Vega) from the 2MASS Extended Source Catalogue. This sample 
covers about $17\%$ of the sky and includes galaxies across $0.003<z<0.3$. 
\citet{Mauch:07} discriminated between AGN and SFGs using emission and 
absorption features in the 6dF spectra and derived local luminosity functions 
for both populations using the  \citet{Saunders:90} form of the luminosity 
function (LF) which is commonly used at long-wavelengths (\ie~mid-IR and 
longward). 
Their sample has also been crossed-matched with the IRAS Faint Source 
Catalogue providing fluxes at 12, 25, 60 and $100\,\um$ which they use to 
confirm the radio-IR correlation. 
The \citet{Mauch:07} catalogue provides us with a local reference sample of
SFGs with known radio, IR and $K-\,$band flux densities (from which we 
can derive SFRs and stellar masses) to which we can compare our results 
from higher redshift.

\section{Radio-based emission diagnostics}
\label{sec.dia}

Here we present four radio emission diagnostics which directly probe the 
physical origin of the radio luminosity: two of these are purely from the 
radio data (radio spectral indices, morphology), and two involve comparisons 
with observed flux densities at other wavelengths (mid-IR and near-IR) as a 
function of redshift. These diagnostics are mostly positive AGN 
discriminators, i.e. they generally imply that the radio emission of a 
source is very likely powered by AGN emission and not star formation. 
We do not include the obvious radio luminosity discriminator, where
a radio luminosity $>10^{25}\,$WHz$^{-1}$ would imply an unphysical 
SFR of $\sim5000\,M_\odot$yr$^{-1}$, as those eight sources clearly above 
this luminosity are flagged as AGN by one or more of the other methods
below and potentially such objects do exist in small numbers at high 
redshift.

We give the number of sources from our parent sample 
for which each of our AGN/SFG discriminators is applicable in 
Table~\ref{tab.num} along with a short comment of why they apply to only 
such a number. We discuss the implications of these selection functions in 
\S\ref{sec:res}.

\begin{table}
 \begin{minipage}{8.5cm}
   \caption{Number of sources from our parent sample for which each 
     SFG/AGN discrimination method is applicable. We list discrimination
     method, the number of sources which it is applicable to and the reason why.}
  \centering
  \begin{tabular}{@{}ccc@{}}
  \hline
   Discrimination & \# & Limitation of method \\
  \hline
   morphology   & 127 & brightest sources only and coverage\\
   spectral index     & 45 & brightest sources only \\
   $S_{\rm 24\um}/S_{\rm 1.4GHz}$  & 445 & limited by $24\,\um$ coverage\\
   $S_{\rm 1.4GHz}/S_{\rm 2.2\um}$ & 449 & applies to all sources\\
   \hline
\label{tab.num}
\end{tabular}
\end{minipage}
\end{table}

\subsection{Radio morphology}

Only six sources from the parent sample show the classical unambiguous AGN 
signature of a double lobed, FR2 morphology \citep[\ie][]{Fanaroff:74} in 
our $3.3$\,arcsec resolution imaging. For other 
extended VLA sources there is not enough detail to distinguish between a 
clear AGN morphology and extended, galaxy scale emission from 
star formation.

Additionally we include results of the combined MERLIN+VLA(A-array) maps at
1.4\,GHz (A. Zoghbi et al., in prep.) which have a resolution of 
$0.5$\,arcsec. These results are limited to the brighter objects from our 
master sample due to 
the sensitivity of our MERLIN images, and also the inner $\sim20\,$arcmins
covered by the four MERLIN pointings. We are therefore able to make combined 
maps for 127 sources. Many show clear AGN jet/lobe morphology including the 
six sources flagged as AGN from the lower resolution VLA images. The other
sources show either a very compact, unresolved morphology or are extended on
$\sim1\,$arcsec scales, often almost completely resolved out
due to the lack of sensitivity of MERLIN to extended structure. Although 
the unresolved sources do not have significantly constrained brightness 
temperatures ($\ge 10^3$\,K) this compactness is most likely due to an AGN 
\citep[\eg][]{Muxlow:05}, but a small chance remains that they could be 
due to nuclear starbursts. The sources with extended emission 
over the galaxy which is often completely resolved out must be from star 
formation occurring throughout the galaxy.  Hence, the MERLIN+VLA images 
show that 42/127 sources are very likely to be AGN. 

\subsection{Radio spectral indices}

Using our $4.8$\,GHz data \citep{Seymour:02} we have radio spectral indices 
for 45 sources in our sample. These are among the brightest sources in our 
sample as the $4.8\,$GHz rms of $30\,\uJy$ corresponds to $\sim70\,\uJy$ at 
1.4\,GHz assuming $\alpha^{4.8}_{1.4}=-0.8$. Radio emission from star 
formation, due to synchrotron radiation, will generally have a spectrum with 
a slope in the range $-0.5\la\alpha^{4.8}_{1.4}\la-1.0$ \citep{Thompson:06}.
Radio sources with spectral indices steeper than $\alpha=-1$
are typically targeted as high redshift radio galaxies. 
Therefore any radio sources with spectral indices {\em not\,} in 
the range $-0.4\la\alpha^{4.8}_{1.4}\la-1.1$ (allowing for the errors in the 
flux densities) we flag as having AGN powered radio emission. We find 31 
sources have spectral indexes not in this range  (\ie~with 
$\alpha^{4.8}_{1.4}\ge-0.4$ or $\alpha^{4.8}_{1.4}\le-1.1$). 

This diagnostic is particularly powerful as it is relatively insensitive to 
redshift; \eg\ \citet{Klamer:06} find that $90\%$ of a sample of high redshift 
AGN have radio spectra best characterised as a power-law with no spectral 
steepening. It is possible that high redshift SFGs could have flatter 
radio spectra due to free-free processes dominating the higher rest-frame 
frequencies, especially if the source is very young. However the 1.4\,GHz 
counterparts to the 4.8\,GHz sources are typically bright ($>1\,$mJy at 
1.4\,GHz) hence unlikely to include many high redshift SFGs. 
Most radio sources classified as AGN by their radio spectral index are also 
classified by other diagnostics (see Table~\ref{tab.break}).

\begin{figure*}
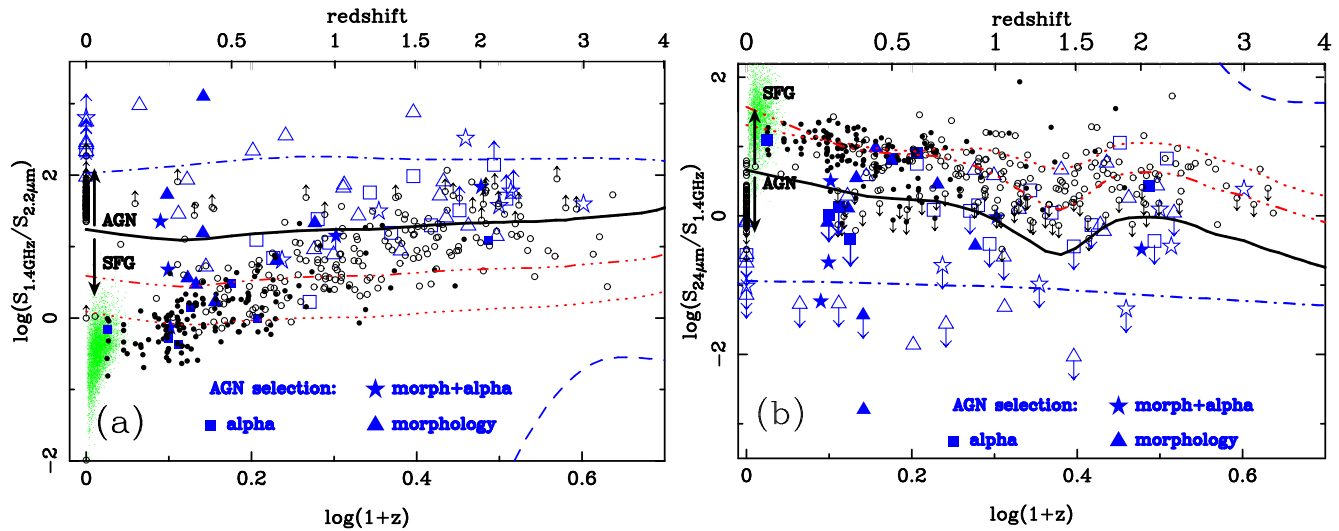

\begin{minipage}{8.8cm}
  \centerline{\hbox{
      \psfig{file=radioloudness.ps,width=8.8cm,angle=270}
    }}
\end{minipage}
\begin{minipage}{8.8cm}
  \centerline{\hbox{
      \psfig{file=rad_ir.ps,width=8.8cm,angle=270}
    }}
\end{minipage}
\caption{{\bf (a)}~1.4\,GHz to near-IR flux density ratio plotted against 
  redshift and {\bf (b)}~$24\,\um$ to 1.4\,GHz flux density ratio plotted 
  against redshift. 
  Filled symbols have spectroscopic redshifts and open symbols have 
  photometric redshifts.  The large blue symbols are those radio sources 
  confirmed as AGN from their morphology (triangles), spectral index (squares) 
  or both (stars).  We show the location of the flux density ratios 
  of the local 6dF-NVSS sample with green dots. The tracks of star forming 
  galaxies with luminosities over the range of interest are marked: LIRG (red 
  dotted lines), ULIRG (red dot-dashed lines). See text more discussion of the 
  templates used. We also over-plot the tracks of a radio-loud (blue dash-dot-dot-dot 
  lines) and a radio-quiet QSO (blue dashed lines) from \citet{Elvis:94}. 
  The cut-off (solid black lines) in both cases is chosen to be 0.6\,dex 
  ($=5\,\sigma$) above or below the more extreme of two 
  the SFG tracks. Sources without known redshifts are plotted at $z=0$.}
\label{fig:radratio}
\end{figure*}

\subsection{Radio to near-IR flux density ratios}
\label{sec.rni}

``Radio-loudness'' is another classical discriminator of radio luminous AGN 
\citep{Kellerman:89} and traditionally based on radio to optical (observed or 
rest-frame) flux density ratio. More 
recently radio-loudness has been defined in terms of absolute radio luminosity 
\citep[\eg][ $L_{\rm 1.4GHz}\ga10^{25}\,$WHz$^{-1}$]{Miller:90}, but this 
diagnostic would only apply to eight of our parent sample and those have already 
been flagged as an AGN by one or both of the previous
two diagnostics. The original definition of radio loudness 
was based on $B$-band observations, but given that at the redshifts in 
question, $0\le z\le 4$, this band 
quickly starts sampling below the $4000\,$\AA\ break where variations in 
star formation rate and absorption by dust have a strong effect, we chose to 
use a longer wavelength band. In fact the only band short-ward of the 
mid-IR which remains longward of the $4000\,$\AA\ break up to $z\sim4$ 
is $K-\,$band. Our $K-\,$band data are particularly deep: $K=22.6$ 
($3\,\sigma$ limit). 

In keeping with our philosophy stated in the introduction, where we make no 
assumptions about the observed SED, we plot the observed flux density ratios 
against redshift. Given that we do not know the nature of each radio source 
beforehand, we can only plot the un-k-corrected flux density ratios and 
compare them with tracks of different AGN and starburst galaxies. This 
approach avoids making any {\it a priori} assumptions about the nature of 
each source. In Figure~\ref{fig:radratio} (a) we present the $1.4\,$GHz to 
$K-\,$band flux density ratio against redshift. We plot the tracks of 
two star forming galaxies which have no strong AGN component. These are 
LIRG and ULIRG templates made from average composites from fits to about 
a dozen LIRGs and ULIRGs with very high fidelity observations. From $5-35\,\um$ 
they are based on {\it Spitzer} IRS long low spectra. 
For $1-5\,\um$, the templates are 
built from stellar population models, with overall slopes constrained by large 
beam IRAC and 2MASS photometry. The radio regime was determined with data from 
\citet{Condon:91}. The corresponding SFRs of these two templates are 
approximately $\sim10$ and $\sim100\,M_\odot$yr$^{-1}$ respectively. We also 
plot the tracks of classical `radio-loud' and `radio-quiet' QSOs
from the templates of \citet{Elvis:94}.

Figure~\ref{fig:radratio} (a) shows several clear radio-loud AGNs with 
flux density ratios similar to or greater than the radio-loud QSO template 
track. We can use this plot to define a locus where a radio source will 
unambiguously identified as an AGN. We decide to take a cut $5\,\sigma$ above 
the higher SFG track (the ULIRG one) where $\sigma$ comes from taking an uncertainty
of $20\%$ in the radio flux density and $20\%$ in the $K-$band magnitude.
We emphasise that these uncertainties are extremely conservative and that the
majority of radio sources will have much smaller uncertainties, but we are 
also allowing for the uncertainty in the SEDs used for the template tracks.
Combining these uncertainties we get a $5\,\sigma$ cut of 0.6\,dex 
($=5\times\sqrt(0.2^2+0.2^2)/ln(10)$).

We note that all bar one of the sources with unknown redshifts are well above
our AGN/SFG discrimination line at almost all redshifts and hence are most
likely all obscured AGN at high redshift (see \S~\ref{sec:noop} for further 
discussion of these sources).

\subsection{Radio to mid-IR flux density ratios}

The correlation of the radio and far-IR luminosities of star forming galaxies
is well established  over five orders of magnitude \citep{Yun:01} and is one 
of the tightest results in astrophysics. The continuation of this relationship 
in the mid-IR and to $z\ge1$ is also fairly well established 
\citep{Garrett:02, Appleton:04}. Furthermore we show in the Appendix 
that the total-IR 
to radio luminosity relationship does continue to high redshifts for ULIRGs. 
This relationship can be used as another clear discriminant of 
radio loud AGN; all objects with a relative ``radio-excess'' are very likely 
to be AGN.

{\it Spitzer} provides comparably deep IR observations to the radio which 
allows comparison of the IR properties of the sub-mJy radio population. We 
choose to use the $24\,\um$ band of the MIPS instrument as it is more 
sensitive than the $70\,\um$ band with a $3\sigma$ detection limit of 
$98\uJy$ compared to $3\sigma\sim3\,\mJy$ at $70\,\um$. This $24\,\um$ 
detection limit should be sufficient to detect all the luminous starburst 
galaxies ($L_{\rm 24\um}\ga3\times10^{10}\,L_\odot$) which we could detect at 
radio wavelengths out to $z\sim1$ and most starburst galaxies at $z>1$, given 
the nominal 1.4\,GHz detection limit of our parent sample.

In Figure~\ref{fig:radratio} (b) we plot the distribution of mid-IR to radio 
flux density ratios, $S_{24\um}/S_{1.4{\rm GHz}}$, against redshift for sources 
in our sample. We compare this distribution to redshifted starburst and AGN 
templates tracks in a similar way to section~\ref{sec.rni}. On the whole a 
large number 
of $24\,\um$ detected sources follow the SFG template tracks suggesting that 
star forming galaxies make up a significant fraction of the sub-mJy radio 
population. There are many obvious ``radio-excess'' and a few 
``IR-excess'' sources which are both likely to harbour AGN. The radio-excess
sources are likely to host AGN due to their radio-loud nature. The IR-excess 
sources are likely to host radio-quiet AGN with hot dust dominating in the 
mid-IR, \ie\ radio-quiet obscured AGN such as those found by \citet{Lacy:07}, 
but with moderate amounts of star formation dominating the radio emission.
These ``IR-excess'' sources likely explain the high the mid-IR to radio flux 
density ratio seen in \citet{Boyle:07} due to the selection effects of that 
study. The general trend toward lower mid-IR ratios at higher redshifts 
and the starburst SEDs presented suggest that the low mid-IR ratio found by
\citet{Beswick:08} is due to k-correction effects to the observed ratio.

As in the previous section we use a $5\,\sigma$ cut-off below the more extreme
starburst. We again get a cutoff value of $5\,\sigma=0.6\,$dex from  
assuming a median $20\%$ uncertainty in both the 1.4\,GHz and
 $24\,\um$ flux density. This cut also allows for 
uncertainty in the SED tracks. 

\subsection{Results of AGN/SFG discrimination}
\label{sec:res}

\begin{table}
 \begin{minipage}{8.5cm}
  \caption{Results of the discrimination of sub-mJy radio sources. The raw 
  numbers are presented for each galaxy type as well the number of with 
  spectroscopic, photometric or unknown redshifts. We also 
  present the number of
  different sources corrected for incompleteness in the radio survey and the 
  percentage of the total number of sources.}
  \centering
  \begin{tabular}{@{}cccccc@{}}
  \hline
   Classification     & N  & N$_{\rm spec}$/N$_{\rm phot}$/N$_?$ & N$_{\rm corr}$ & Percentage \\
  \hline
   AGN   & 178 & 31/88/59  & 244 & $35.8\%$\\
   SFGs  & 269 & 131/138/0 & 436 & $64.0\%$\\
   Stars & 1   & 1/0/0     & 1   & $0.1\%$\\
  \hline
\label{tab.class}
\end{tabular}
\end{minipage}
\end{table}

Using the above four discriminators we can discriminate the AGN from the SFGs
on a statistical basis. Although most of these discriminators are mainly just 
positive AGN identifiers we can be reasonably confident that we have removed 
nearly all the AGN, at least statistically. In fact, $58\%$ of the sources 
flagged as AGN are done so by two or more discriminators.
We find that 178/449 sources show clear indications that their radio emission
is due to AGN activity. One radio source is, surprisingly, identified as a 
star from optical spectroscopy.
These results are summarised in Table~\ref{tab.class}, along with the 
incompleteness corrected number and percentage. This incompleteness factor is
a correction for the decrease in the sky area probed at faint fluxes due to 
the decrease in sensitivity of our radio map away from the pointing centre.

\begin{table}
 \begin{minipage}{8.5cm}
  \caption{Break-down of the number of radio sources classified as AGN by 
    each of our four discrimination methods. 
    The columns and rows correspond to the four methods of AGN/SFG 
    discrimination used in sections 3.1 to 3.4: radio morphology, radio 
    spectral index, mid-IR/radio flux density ratio and 
    radio/near-IR flux density ratio. The diagonal represents the number of 
    sources classified as AGN by just the corresponding discriminator.
    The lower left of the table, below the 
    diagonal, gives the number of sources flagged as AGN for a particular 
    combination of discriminators. }
  \centering
  \begin{tabular}{@{}cccccc@{}}
  \hline
   classification     & morphology & $\alpha$ & mid-IR & near-IR\\
  \hline
  morphology       & 44 & -  &  -   & - \\
  $\alpha$         & 12 & 32 &  -   & - \\
  mid-IR           & 27 & 16 & 102  & -  \\
  near-IR          & 30 & 15 & 79   & 137\\
  \hline
\label{tab.break}
\end{tabular}
\end{minipage}
\end{table}

We look at the break down of radio AGN classification in Table~\ref{tab.break}. 
We find that the two most effective methods are the flux density ratios 
which each account for $\sim60-80\%$ of the sources classified as AGN, but 
only overlap for $\sim44\%$ of the total. Hence, 
together they account for almost all, $90\%$ (160/178), of the radio sources 
classed as AGN. The sensitivity of our $K-\,$band and $24\,\um$ observations 
are generally deep enough that we are not biased by 
non-detections in these bands as that would put a source clearly out of the 
SFG regime, except possibly for the mid-IR/radio ratio where above redshift 
1 we get a few $24\,\um$ non-detections in the SFG regime for the very faintest
radio sources. The morphology and spectral index  methods are 
not as powerful (see table~\ref{tab.num}), being primarily limited by the 
sensitivity and coverage of the current observations. Of the sources 
classed as AGN, these discriminators are together responsible for 
classifying $35.9\%$, far less than that due to the flux ratio cuts, but 
these two other selection criteria tend to agree with the flux ratio 
methods.

We also investigated the effect of systematically varying the two main 
discriminators, the flux density ratios plotted against redshift, by 
$\pm1\,\sigma$ ($=0.12\,$dex). When both cuts were changed so as to classify 
more radio sources as SFGs we found a swing (\ie\ change in AGN/SFG 
distribution) of $4.2\%$ in the number of SFGs from the parent sample and when 
changed so as to increase the number of AGN we found a swing of $5.6\%$. These 
values are on par with the Poisson statistics expected from dividing the 
parent sample into two roughly equal populations ($\sim6\%$).
It is possible that some or many of the SFGs may contain AGN, but that these 
AGN do not contribute strongly to the radio emission, \eg~ the ``IR-excess'' 
sources in Figure~\ref{fig:radratio}(b). 
It is also possible that some radio SFG are really radio AGN, especially 
as some of the discrimination methods do not work so well at faint flux 
densities. The possible number of AGN interlopers is difficult to quantify 
other than varying the discrimination criteria (see previous paragraph) 
without superior data.

\subsubsection{Radio sources with faint optical or near-IR detections}
\label{sec:noop}

\bfig
\psfig{file=flux_rmag2.ps,width=8.4cm,angle=270}
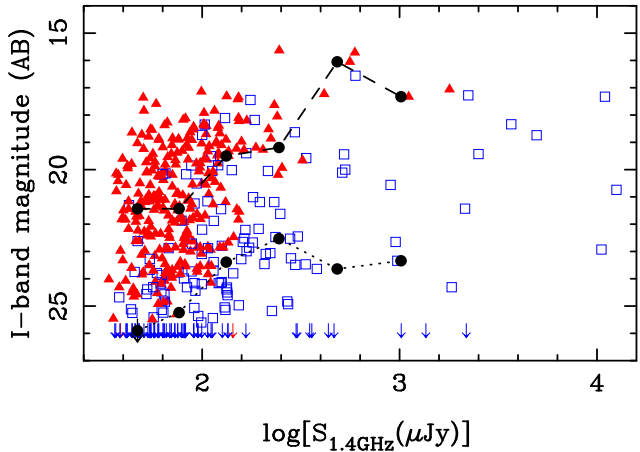
\caption{The 1.4\,GHz radio flux density plotted against $I-$band magnitude of 
  all our sources separated into AGN (blue squares) and SFGs (red triangles). 
  Radio sources undetected in $I-$band above $3\sigma$ are marked as upper 
  limits at $I\sim 25.9$. The lines represent the median $I-$band magnitude 
  for the SFGs (dashed line) and AGN (dotted line) as a function of flux 
  density. }
\label{fig:radmag}
\efig

Determining the nature of faint radio sources with weak or no detections at 
other wavelengths is naturally difficult. 
We have 60 sources from the parent sample where we are not 
able to determine a redshift, spectroscopic or photometric. In fact, 31 of 
these sources have no detections, $>3\,\sigma$, at any other 
wavelength. The radio sources with unknown redshifts are included in 
Figure~\ref{fig:radratio} (both panels) at $z=0$, either as upper/lower 
limits or detections. Their positions in these plots suggest that they 
would be mainly classified as AGN whatever their redshift, especially in 
Figure~\ref{fig:radratio}(a), where our AGN/SFG cutoff has only a weak 
dependence on redshift. We considered the possibility that they are low mass 
SFGs at $z\sim1$, \ie\ not detected in our deep $K-$band of IRAC data. 
Such an object would have a SFR of $\sim100\,M_\odot$yr$^{-1}$, but a stellar 
mass $\la 10^{8.5}\,M_\odot$. A few sources with characteristics close to 
these values have been seen, \eg\ the hosts of GRBs, but the SFR and 
non-detection in our $B-$band observations implies UV extinction of 
$A_{\rm V}\sim8$ over the whole galaxy. While we may have 
missed a couple of extreme star forming sources it is unlikely that many of 
these sources are starbursts at moderate redshift.

We present the radio flux density verses $I-$band magnitude distribution in 
Figure~\ref{fig:radmag}. This plot shows that the brightest radio sources 
($\log(S_{\rm 1.4GHz}/\mu Jy)>2.5$) are mainly AGN, but with a population of 
star forming galaxies coming in at lower radio flux densities and brighter 
optical magnitudes than the AGN. This result also suggests that the optically 
un-identified and faint sources are most likely to be AGN (\eg\ at high 
redshift and obscured). At $z\ge 1$ sources
would appear compact enough that we should detect intrinsically bright, but 
low surface brightness galaxies at the depths of our data which have a typical 
seeing of $1\,$arcsec. Below $z=2.2$ we expect to detect all sources with 
$\log(L_{\rm 2.2\um}/L_\odot)>10$, 
in our near-IR or IRAC data, corresponding to quite low stellar masses
(i.e. $\log(M/M_\odot)\sim 10$ depending on the exact mass-to-light ratio). 
We plot the median $I-$band magnitude of the SFGs and AGN in 
Figure~\ref{fig:radmag} which both decrease toward fainter radio flux 
densities.  We believe the parent sample to be reliable as we would expect 
0.03 of sources to be spurious in pure Gaussian noise, although the noise in 
radio maps is not completely Gaussian.
In S04 we carefully examined visually all the sources to remove those
that were clearly due to artifacts from the radio reduction. We can't rule 
out the possibility that one or two sources at low S/N and 
with no optical remain spurious, but
our results would not change significantly if this were the case.

Figure~\ref{fig:radmag} shows that the optical properties of the parent 
sample do change with radio flux density contrary to the results 
of \citet{Simpson:06} who do not see a decrease in the median magnitude with 
decreasing radio flux density. The difference between the results here 
and those of \citet{Simpson:06} can mainly be explained by the fact we have 
deeper radio data (by 0.5\,dex) and the trend we see is strongest at 
the faintest flux densities. Results at the highest flux densities are likely 
affected by small number statistics. Hence, we conclude that most, if not all, 
of the radio sources with optically very 
faint counterparts are high redshift radio galaxies (their optical faintness 
implying they are obscured, type 2 AGN). They are likely to be similar to 
the classical high redshift radio galaxies, but less luminous \citep[\ie\ in 
the $24\le \log(L_{\rm 1.4GHz}/$WHz$^{-1})\le26$ range][]{Sajina:07}.

\section{Radio source counts by type}
\label{sec.cnts}

The total Euclidean normalised radio source counts from the $13^{\rm H}$ field 
were presented in S04 where we found the counts to be in 
agreement with the many other published counts at similar flux densities. 
However, a scatter between the different counts from different surveys was 
found which was particularly strong around $200-300\,\uJy$ with the counts 
from the HDF-N \citep{Richards:00} being significantly low. This scatter was 
attributed to sample variation due to large scale structure. An independent 
reanalysis of the HDF-N counts by \citet{Biggs:06} revised them upwards, but a 
smaller scatter due to sample variance nevertheless remains. However, we note 
that the counts from the 
$13^{\rm H}$ field, Phoenix Deep Field Survey \citep{Hopkins:03}, HDF-S 
\citep{Huynh:05} and the three fields from \citet{Biggs:06} are consistent 
with only a $10-20\%$ scatter over the $70-500\,\uJy$ flux density range.

We can now re-calculate the source counts at 1.4\,GHz separately by type, AGN 
and SFG, correcting for incompleteness of the radio survey as in S04. As in 
the original presentation of these counts we only take sources above 
$5\,\sigma$. Uncertainties include Poisson statistics and sample 
variance (which we discuss in \S~\ref{sec.sfh}). 

\begin{figure}
  \centerline{\psfig{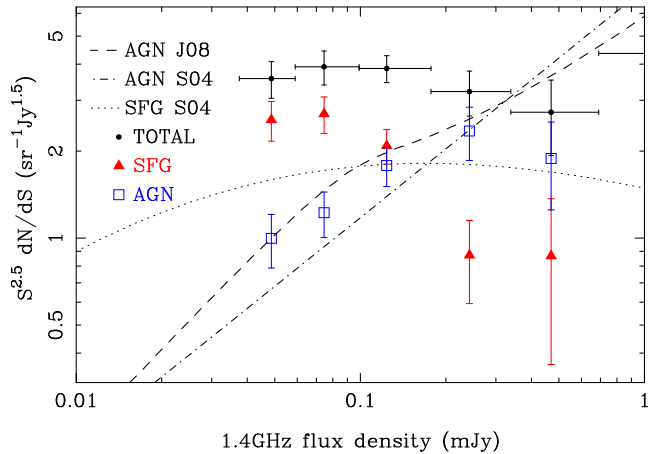}}
    \caption{The Euclidean normalised 1.4\,GHz source counts for the 
      $13^{\rm H}$ deep survey: total counts (filed circles), SFGs (triangles) 
      and AGN (open squares). Error bars represent uncertainties derived from 
      Poisson statistics and sample variance. We overlay AGN and SFG models 
      from S04 and M. Jarvis et al. (in prep. J08). Note that a large part of 
      the up-turn in counts above the extrapolation of earlier AGN models 
      (S04 etc.) from brighter 
      flux densities is explained by SFGs, but that the AGN population partly 
      contributes (about one quarter) to the up-turn even in the faintest bins.}
\label{fig:counts}
\end{figure}

The counts by type are presented in Table~\ref{tab.counts} 
and Figure~\ref{fig:counts}. We use slightly different 
binning than in S04 to obtain the most uniform distribution of numbers 
for both AGN and SFG populations per bin. We find that there is a rapid rise 
in the contribution to the counts from SFGs as expected from modeling of the 
counts \citep{RowanRobinson:93,Hopkins:98, Hopkins:04, Seymour:04, Huynh:05}.
The star forming galaxies are the more numerous population below $120\,\mu$Jy 
although the AGN still contribute approximately one quarter of the total 
counts even in the faintest bin. So while SFGs dominate, AGN do have a small 
contribution to the up-turn, but maybe not as much as suggested by 
\citet{Simpson:06}. The fraction of SFGs in the two lowest flux density bins, 
$72\pm22\%$, is  consistent with the $\sim70\%$ of $<0.1\,$mJy sources found 
to be SFGs by \citet{Muxlow:05} in high resolution ($0.2-0.5$\,arcsec) 
MERLIN+VLA imaging of the HDF-N by. We note that no sources were detected 
between $27-40\,\uJy$ (S/N$>8\sigma$) in the inner $3.4\times3.8$\,arcmin in 
that work. Those authors inferred that the $\sim8$ sources expected from the 
extrapolation of the source counts were resolved out on scales $>0.5$\,arcsec 
and hence, in our opinion, are likely to be mostly SFGs, \ie\ a SFG fraction 
close to $100\%$.

The sharp disconnect in the number of SFGs from 0.2 to $0.1\,$mJy is perhaps 
due to the fact that our AGN selection method may be more efficient at higher 
flux densities, but it most likely represents the sharp rise in the luminosity 
of a typical SFG from $z=0$ to $z=1$ and the small number statistics in the 
highest flux density bins. The effect of slightly varying the flux 
ratio cuts in the two main AGN/SFG discrimination methods (as described in 
\S~\ref{sec:res}) does not qualitatively change 
the results in Figure~\ref{fig:counts}. When changing the cut-offs in flux 
ratio by $0.12\,$dex in favour of increased SFG numbers the swing in the 
faintest flux density bin is $5.2\%$ and 
when changing them in favour of the AGN the swing in this bin $10.4\%$. These 
swings are less than the quoted uncertainties on the source
counts in this bin, $\sim20\%$.

We overlay some models for the AGN and SFG contribution to the radio source 
counts. We show the SFG and AGN model from S04 and the AGN model courtesy of 
M. Jarvis and R. Wilman, based on \citet{Jarvis:04}. The S04 AGN model is 
basically an extrapolation from higher flux densities of the models used in 
\citet{Dunlop:90} and \citet{Hopkins:98} and hence not necessarily valid for 
these low flux densities although they were the best available at the time. 
The Jarvis AGN model matches the observed AGN counts well and includes a 
component of radio quiet QSOs based on the evolution of the X-ray AGN 
luminosity function. This model appears to follow reasonably well the increase 
in AGN above the extrapolation of older models, like that used in S04, from 
higher flux densities. The S04 SFG model is derived from the local luminosity 
function of SFGs \citep[from][]{Sadler:02} with luminosity evolution of the 
form $(1+z)^{2.5}$ to $z=2$ (and constant thereafter), and does not fit the 
source counts well by over-predicting the SFG counts at the bright end 
and under-predicting them at the faint end. The discrepancy at the bright end 
may be due to the small volume probed at low-redshift where the brightest SFGs
are likely to lie. 

\begin{table}
 \begin{minipage}{8.4cm}
  \caption{Euclidean Normalised 1.4\,GHz Source Counts: Total and by Galaxy 
    Type. The first column shows the range of flux densities of each bin. The
    second column is the middle of the bin in log space. The last three 
    columns indicate the Euclidean normalised source counts: total and for 
    AGN and SFGs separately for the first time with their associated 
    uncertainties.} 
  \centering
  \begin{tabular}{@{}ccccc@{}}
  \hline
   Flux Range & Flux      &        & Counts & \\
   ($\mu$Jy)  & ($\mu$Jy) &    & (sr$^{-1}$Jy$^{1.5}$) & \\
              &           & total  & AGN & SFGs \\
  \hline
$  37.5 -  55.7 $ &  45.7 & 3.48 $\pm$ 0.53 & 0.98 $\pm$ 0.22 & 2.50 $\pm$ 0.41 \\  
$  55.7 -  89.6 $ &  70.7 & 3.94 $\pm$ 0.54 & 1.11 $\pm$ 0.21 & 2.83 $\pm$ 0.41 \\ 
$  89.6 - 156.1 $ & 118.3 & 4.12 $\pm$ 0.42 & 1.61 $\pm$ 0.24 & 2.51 $\pm$ 0.32 \\  
$ 156.1 - 294.1 $ & 214.2 & 2.73 $\pm$ 0.46 & 1.82 $\pm$ 0.37 & 0.91 $\pm$ 0.27 \\  
$ 294.1 - 600.0 $ & 420.1 & 3.46 $\pm$ 0.79 & 2.53 $\pm$ 0.64 & 0.99 $\pm$ 0.50 \\  
\hline
\label{tab.counts}
\end{tabular}
\end{minipage}
\end{table}

\section{Properties of the radio star forming galaxy population}

\subsection{Star Formation Rates and Stellar Masses}

\begin{figure*}
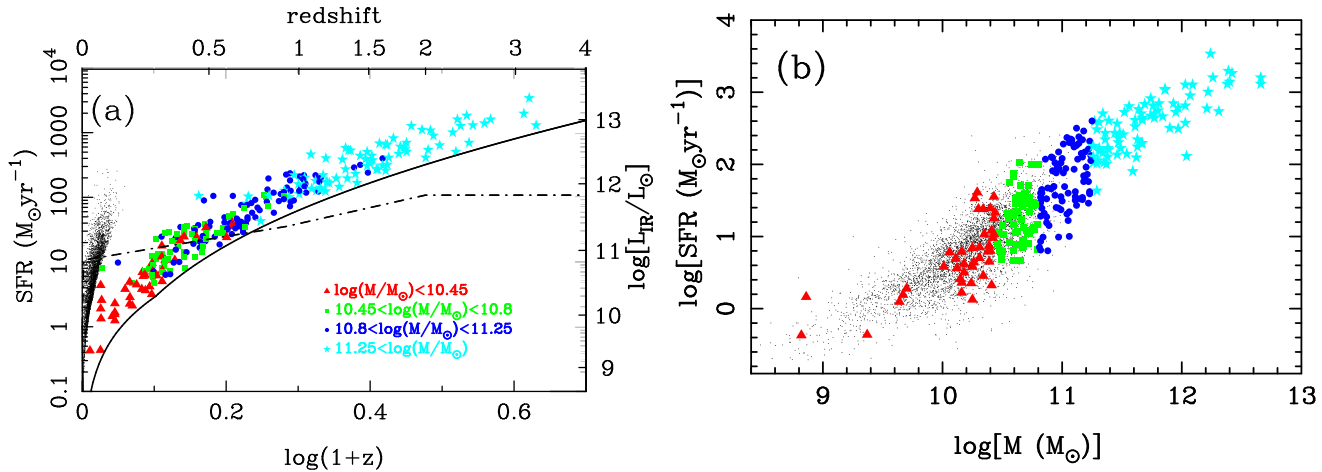

\begin{minipage}{8.7cm}
  \centerline{\hbox{
      \psfig{file=zlsfr.ps,width=8.6cm,angle=270}
    }}
\end{minipage}
\begin{minipage}{8.7cm}
  \centerline{\hbox{
      \psfig{file=mvsfr.ps,width=8.6cm,angle=270}
    }}
\end{minipage}
\caption{SFRs of the radio SFGs from the $13^{\rm H}$ field plotted 
  against redshift {\bf (a)} and stellar mass {\bf (b)}. 
  The different symbols represent different stellar masses 
  ranges as indicated in panel (a). In panel (a), the solid 
  line represents the 1.4\,GHz detection limit and the dot-dashed line 
  indicates the 10 times $L_\star$ of 
  the radio star forming luminosity function taking a luminosity evolution of
  $Q=2.5$ which we use in Section~\ref{sec.mvsfr}. The black 
  dots are the star forming galaxies from the local 6dF-NVSS sample.}
\label{fig:zsfr}
\end{figure*}

The conversion of radio luminosity to SFR \citep{Bell:03} relies on the well 
known radio-IR relation \citep[\eg][]{Yun:01}. While we show in the Appendix 
that the radio-total IR relation does indeed hold to high redshifts/high 
luminosities, to derive SFRs from the radio/IR correlation in such 
situations requires two subtle corrections. The first is that in determining 
the local correlation,  
\citet{Yun:01} did not apply k-corrections, which become significant for the 
highest luminosity galaxies in their sample. 
The second issue is that some of these 
galaxies have relatively flat radio spectra, so correcting those observed at 
high redshift back to 1.4\,GHz using a nominal slope of $-0.7$ does not 
provide a good estimate. When nominal k-corrections are 
applied to the Yun et al. sample, the best fit ULIRG ratio of 
$L_{\rm 60\um}/L_{\rm 1.4GHz}$ is $\sim140$. 
We have used the observations of Condon et al. 
(1991) to determine a more accurate slope. We find a final ratio of 
$L_{\rm 60\um}/L_{\rm 1.4GHz} = 128$ for a galaxy at $z\sim2$ observed at 1.4\,GHz 
and corrected as if it had a slope of $-0.7$.  This result would imply that 
the SFRs estimated above using the standard local ratio are under-estimated 
by about 0.1\,dex, a correction we have applied in our analysis. After these 
corrections and our choice of a \citet{Kroupa:01} IMF, 
our conversion from radio luminosity to SFR is $0.07\,$dex less than that
of Bell (2003), \ie\ for a given radio luminosity we
assume a SFR 0.84 times that predicted by Bell (2003).

We plot the SFRs of our SFG sample against redshift in Figure~\ref{fig:zsfr}(a) 
using different symbols for objects of different stellar masses. 
Stellar mass estimates are calculated for the $13^{\rm H}$ sample of SFGs 
by normalising a M82 SED to our IRAC $3.6\,\um$ fluxes ($K-$band for 8 sources
not covered by the {\it Spitzer} data) and using a \citet{Kroupa:01} IMF 
to determine a rest-frame 
$H-$band luminosities and assuming a M82 mass-to-($H-$band)-light ratio. 
Similarly, we derived stellar masses for the 6df-NVSS sample from their 
observed $K-$band magnitudes assuming the same M82 SED and mass-to-light-ratio.
Clearly using one mass-to-light ratio for a sample of
galaxies with a range of SFRs, amongst other properties, is not ideal, but 
this approach suffices for the current investigation. The possible selection 
effects of this choice are discussed in section~\ref{sec.se}. 

We find two striking results. Firstly, there are many very high SFR, 
$>300\,M_\odot$yr$^{-1}$, sources above $z=1$ 
compared to the local Universe. The 6dF-NVSS Survey covers a large area of the 
sky, $17\%$ equivalent to $\sim7\,$Gpc$^3$ to $z=0.3$, hence 
is not biased against detecting rare, exceptionally high SFR sources. Therefore, 
the lack of exceptionally high SFR sources locally  is not a volume 
selection effect and the presence of very high SFR galaxies at high redshift 
is observed in surveys  at other wavelengths 
\citep[\eg\ in the sub-mm][]{Chapman:05,Muxlow:05,Pope:06}. 
Secondly, all our high SFR galaxies have very high 
stellar masses. We discuss this result more within the framework of current
galaxy evolution models in Section~\ref{sec.mvsfr}. 

We can compare the SFR against stellar mass for both our sample of SFGs
and the 6df-NVSS sample in Figure~\ref{fig:zsfr}(b) and we find that a trend 
of SFR correlating with stellar mass is apparent in both samples (\ie\ from 
the local Universe up to $z\la 3$). 
As radio flux limited surveys will always sample the highest SFR objects
at any given redshift, this result is consistent 
with the idea that there is an upper limit to the SFR of a galaxy of a given 
mass. We see a similar trend in the results of \citet{Noeske:07a} who 
find a ``main sequence'' relation between SFR and stellar mass at a given 
redshift with increasing SFR observed at higher stellar masses. 
The hypothesis that lower mass galaxies can only achieve lower
maximum SFRs could be explained by the fact that more massive galaxies
generally have more baryons in gas and dust which can act as raw material for 
the star formation. 
Potentially some of the very high radio luminosities could be over-estimated
due to uncertainties in both the photometric redshift and k-correction (\ie\
radio spectral index).
Our radio/near-IR flux ratio cut does affect the number of sources detected 
with high SFR to mass ratio, at high masses/SFR/redshift. However, we have 
demonstrated in Section~\ref{sec:res} that by varying this cut by 
$1\,\sigma=\pm0.12\,$dex does not affect the results our 
AGN/SFG discrimination very much. We may have one or two AGN interlopers, but 
we are likely complete in a statistical sense.

\subsection{Star formation history of the Universe across $0<z<3$}
\label{sec.sfh}

The star forming galaxies are of great interest as their radio luminosity 
provides a relatively {\em unbiased} measure of their star formation rate 
independent of the effects of dust. Using this tracer of star formation we can 
probe the global cosmic star formation rate \citep[\eg ][]{Lilly:96, Madau:96}.
We restricted the sample to all sources with $z\le 3$ as we only 
detect two SFGs above $z=3$.

The star formation rate densities are calculated using the standard 
$1/V_{max}$ method which allows for the change in detection limit across the 
relatively broad 
redshift bins. The redshift bins are chosen to be equal in size in $\log(1+z)$ 
space and to have at least 50 sources per bin apart from the highest redshift 
bin which has 13 sources. The 
redshift bins are deliberately left relatively wide to mitigate large scale 
structure/sample variance, Poisson statistics and any possible systematics in 
the uncertainties of the photometric redshifts. The total
star formation rate density in each redshift bin requires two corrections:

\begin{itemize}
\item {\it Radio survey incompleteness:} 
  The radio survey is not complete down to the $30\,\uJy$ detection limit at 
  1.4\,GHz, mainly due to the attenuation of the primary beam of the VLA away 
  from the pointing center, but also due to other instrumental effects such as 
  band-width smearing. Hence the radio sources at low signal-noise, 
  $\la100\,\uJy$, need to be corrected by a weighting factor representing the
  sources not detected in our survey. The weighting
  factor for each source is a measure of the sky area sampled by the
  1.4\,GHz survey at the flux density of the source.
  This correction factor is a function of flux density and position away 
  from the pointing center, i.e our detection 
  limit is $\sim60\,\uJy$ at the edge of the field. This 
  correction is described in full in S04 and we apply the 
  individual weightings from that work to each source.

\item {\it Star forming radio sources below the nominal 1.4GHz VLA 
  detection limit:} As shown in panel (a) of Figure~\ref{fig:zsfr}, our 
  survey primarily
  probes only the high luminosity end of the the SFG luminosity function, 
  especially at high redshift. Any estimate of the star formation rate 
  density needs to include the contribution from sources further down the 
  luminosity function and below the detection limit. Clearly the luminosity 
  function must evolve rapidly given the fits to the source counts 
  \citep[][etc.]{Hopkins:98,Seymour:04} and the appearance of very high SFR 
  sources at $z>1$.

  The evolution of the SFG luminosity function remains slightly uncertain, 
  but most results find a luminosity evolution of $Q=2-3$, using the form 
  $(1+z)^Q$, with negligible density 
  evolution \citep{RowanRobinson:93,Seymour:04,Hopkins:04,Huynh:05,Moss:07}.
  \citet{Haarsma:00} use a different functional form of the evolution with 
  more free parameters, but find quantitatively the same behaviours for 
  luminosity and density evolution as the authors above. However, 
  \citet{Haarsma:00} do not provide uncertainties for their fitted parameters.
  As there is no consensus on the evolution parameter in the literature, 
  or the exact form of the evolution,  we take a representative value of 
  $Q=2.5\pm0.5$,  which is consistent with preliminary results from fitting 
  the radio LF (D04), and $P=0$ for the
  luminosity and density evolution parameters respectively.  We then use the 
  local radio SFG LF of \citet{Mauch:07}. We use these evolution
  parameters to calculate the fraction of the luminosity density below 
  our nominal detection limit for each redshift bin. We give the 
  corresponding multiplicative correction factors and associated 
  uncertainties in table~\ref{tab.sfrd}.
\end{itemize}

\begin{figure}
  \centerline{\psfig{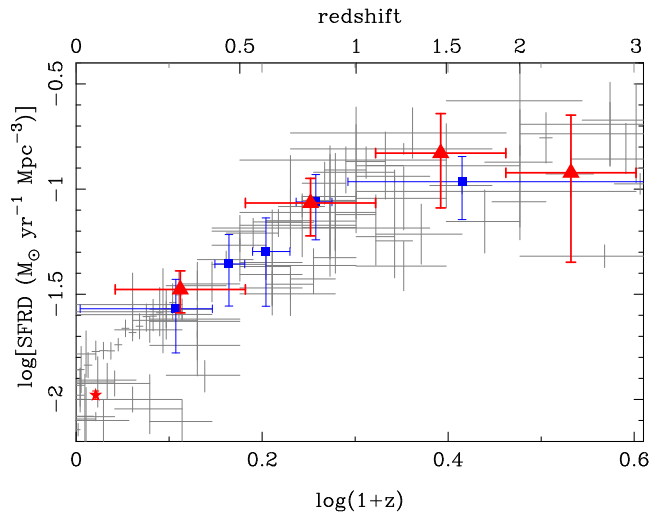}}
  \caption{The comoving star formation rate density (SFRD) of the Universe as 
    a function of redshift from this work (red triangles) and other surveys. 
    The horizontal error-bars represent the width of the bins and 
    vertical error-bars represent the combined uncertainties 
    (see text for details). The SFRD from a 
    selection of other methods (UV, $H_\alpha$, far-IR etc.) taken from a 
    compilation in \citet{Hopkins:04}, are marked 
    by gray lines indicating bin widths and associated uncertainties. 
    From this sample we have highlighted (as blue squares) the results of 
    \citet{Haarsma:00} which are also derived from radio observations.
    The local ($z=0.05$) value of the radio derived SFRD from \citep{Mauch:07} 
    is plotted as a red star with uncertainties smaller than the symbol 
    size. }
\label{fig:sfrd}
\end{figure}

The uncertainties include the following effects:

\begin{itemize}
\item {\it Poisson uncertainties:} We chose the redshift bin size to include at 
  least $\sim50$ sources per bin apart from the highest redshift bin.
\item {\it Uncertainty in the evolution of the Luminosity Function:} 
  As mentioned earlier in this section, the amount of luminosity evolution of 
  the LF remains uncertain so we have used a representative value of 
  $Q=2.5\pm0.5$. Hence we include this uncertainty when derived the uncertainty
  of the SFRD. The effect of including this factor gets stronger at
  higher redshifts as can be seen in Figure~\ref{fig:sfrd} and 
  the LF correction factor in Table~\ref{tab.sfrd}.
\item {\it Sample Variance:} Sometimes referred to as Cosmic Variance, 
  this effect is really quantifying how representative our narrow field 
  is of the whole sky. The source counts 
  between different fields do vary by $10-20\%$ as discussed earlier, but 
  accurately measuring the field to field variation between different surveys 
  is tricky due to different methods used and in some cases different 
  telescopes to correct for instrumental effects that dominate at the faint end.
  This issue needs to be addressed by the different teams working on deep 
  radio surveys although we previously noted in Section~\ref{sec.cnts} that 
  several major surveys are now in reasonable agreement. 
  Hence, to allow for sample variance we simply add a further $20\%$ uncertainty.
\end{itemize}

We further investigate the effects of changing the two main discriminators
(the two flux density ratios as described in Section~\ref{sec:res}).
When changing both these cuts in favour of the SFGs we get 
an increase in the densities, but this is barely noticeable in the lowest 
redshift bin and increases the highest redshift bin by just $10\%$.
When changing the cuts the other way, 
in favour of the AGN, the densities decrease, but only by a few percent at 
the lowest redshift and by $23\%$ in the highest redshift bin. Clearly the 
effect of changing the selection criteria is strongest in the highest redshift
bin, but this result is not unexpected since this bin typically contains the 
lowest S/N radio sources and lowest number of sources. The flux ratio 
discriminators are least good at the highest redshifts, and the radio 
morphology/spectral index methods cannot be applied because most objects are 
too faint. Furthermore, these changes are generally smaller than 
the combined uncertainties from the effects listed above and hence indicate
that the AGN/SFG discrimination criteria used in Section~\ref{sec.dia} are 
not the largest uncertainty in deriving the SFRD at the highest redshift.

\begin{table}
 \centering
  \caption{Star Formation Rate Density at different Epochs. The first column 
    represent the size of the redshift bin. The second column indicates the 
    middle of the bin in $\log(1+z)$. The third column gives the number of 
    SFGs in each redshift bin. The fourth column indicates the multiplicative 
    correction due to the part of the LF below the detection limit using the 
    SFG LF with luminosity evolution of the form
    $(1+z)^{2.5\pm0.5}$. The fifth column presents the comoving 
    star formation rate density derived from the SFGs and the uncertainty.}
  \begin{tabular}{@{}ccccc@{}}
  \hline
   Redshift & $<z>$ & \#& LF cor. & SFRD     \\
   Range   &   &      &   & log(M$_\odot$yr$^{-1}$Mpc$^{-3}$)   \\
  \hline
  0.10 $-$ 0.52 & 0.29 & 104 & $1.35^{+3.5\%}_{-3.8\%}$ &$-1.48^{+0.09}_{-0.11}$  \\
  0.52 $-$ 1.10 & 0.79 &  87 &$2.68^{+21.0\%}_{-20.0\%}$&$-1.07^{+0.12}_{-0.16}$  \\
  1.10 $-$ 1.90 & 1.47 &  54 &$4.60^{+48.3\%}_{-38.1\%}$&$-0.83^{+0.19}_{-0.26}$  \\
  1.90 $-$ 3.00 & 2.40 &  15 &$6.43^{+81.6\%}_{-53.2\%}$&$-0.92^{+0.27}_{-0.43}$  \\
 \hline
\label{tab.sfrd}
\end{tabular}
\end{table}

The results of our determination of the comoving star formation rate density 
of the Universe as a function of redshift are presented in 
Table~\ref{tab.sfrd} and Figure~\ref{fig:sfrd}. These results agree well with 
the multi-wavelength sample from \citet{Hopkins:04},
converted to a Kroupa IMF, 
showing the rapid rise from the local value to a value over an order of 
magnitude larger at $z\ge 1$, followed by a flattening above $z\sim1.5$. We 
highlight the results of \citet{Haarsma:00}, as blue squares corrected to the 
same IMF and cosmology, who also derived SFR comoving density from a deep 
VLA observation of several fields. Their 
results are consistent with ours at all redshifts. \citet{Haarsma:00} perform 
similar corrections for instrumental effects and for sources below the nominal 
detection limit, but have very different AGN/SFG discrimination methods and 
less sophisticated methods of redshift determination (\ie\ for sources without
spectroscopic redshifts those authors use estimates from $I$ or $HK'$ 
magnitudes, or have random assignments). Their uncertainties
are relatively small as they do not include uncertainties in sample variance. 
Our results are also consistent with the results of \citet{Ivison:07b} who use 
stacking of different galaxy types in the radio to derive SFR densities at 
different redshifts. 
The largest uncertainty in our work is that from the evolution of the 
luminosity function, necessary to correct for the un-sampled part of the LF.

\subsection{Contribution to star formation rate density by galaxy stellar mass}
\label{sec.mvsfr}

Recent studies of the star formation history of the Universe have not only 
measured the rapid change in global SFR density, but also the demographics of 
the star forming galaxies. Many authors have observed the evolution of the 
distribution of star formation moving high mass galaxies at high redshift 
to lower mass galaxies at lower redshifts 
\citep[\eg][]{Cowie:96,Panter:04,Juneau:05,Panter:07,Noeske:07a}. 
This effect has been termed ``downsizing'' and is analogous to (and possibly 
related to) the apparent shift of the peak number density of AGN to lower 
redshifts for lower X-ray luminosities \citep[\eg\,][]{Ueda:03}.

\begin{figure}
\begin{minipage}{8.5cm}
  \centerline{\hbox{
      \psfig{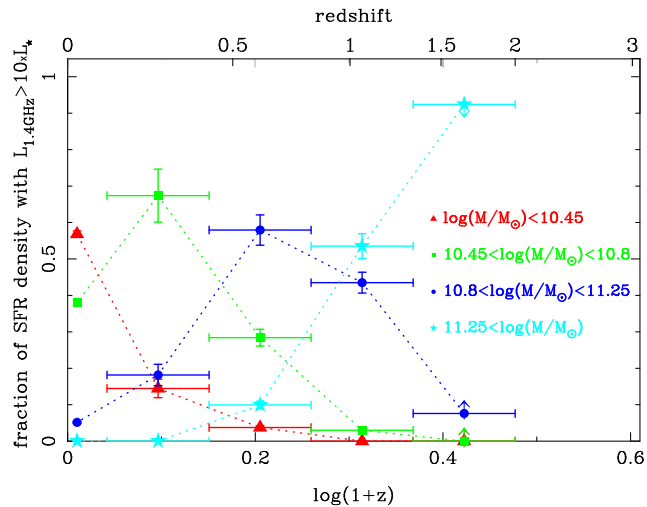}
    }}
\end{minipage}
\caption{The contribution by luminosity/SFR density to the  
  $L_{\rm 1.4GHz}>10\times L_{\star}$ range of the SFG LF. SFGs with different 
  mass ranges are indicated on the plot following our earlier convention. 
  This high end of the luminosity function (in the form from 
  Saunders et al. 1990) represents about $28\%$
  of the SFRD at any particular redshift. The points at $\log(1+z)=0.02$ are 
  from the 6dF-NVSS SFG sample and the other points at higher redshift are 
  from the $13^{\rm H}$ SFG sample. }
\label{fig:sfrdm}
\end{figure}

We are able to examine ``downsizing'' too with our unique radio-selected sample
of star forming galaxies across $0<z<2$. In this section we chose to look just 
at the contribution from the $L_{\rm 1.4GHz}>10\times L_\star$ galaxies where 
we are mostly complete up to $z=2$, Figure~\ref{fig:zsfr}(a). The 
$L_{\rm 1.4GHz}>10\times L_\star$ top end of the luminosity function 
represents the top $28\%$ of the total SFR density at any redshift, for the 
best fitting Mauch and Sadler (2007) LF assuming pure luminosity evolution. 
Hence we can divide each 
SFRD bin into the contribution by stellar mass including that of the local 
sample from \citet{Mauch:07}. We only show the fractional contribution by 
luminosity/SFR density in Figure~\ref{fig:sfrdm}, but note that the 
distribution by number density is very similar. Hence,  
number density and luminosity density are both good tracers of downsizing.

While the sample shown in Figure~\ref{fig:sfrdm} represents about one quarter 
of the star formation occurring at a given cosmic epoch, the 
$L_{\rm 1.4GHz}>10\times L_\star$ sources are the most active at any redshift. 
We find that there is a dramatic change in the stellar mass of the galaxies 
contributing to this part of the luminosity function. The highest redshift bin is 
dominated by massive galaxies, $\log(M/M_\odot)>11.25$, although at this redshift 
we are not quite complete to $L_{\rm 1.4GHz}=10\times L_{\star}$ and hence we mark 
points in this bin as limits. In the lowest redshift bin (from the 6df sample) we 
can see the dramatic, rapid rise in the number of 
the least massive galaxies. We are even able to see the rise and fall of the 
contribution from intermediate stellar mass ranges at successive intermediate 
redshift ranges. 

These results are qualitatively similar to other recent results 
\citep[e.g.][]{Juneau:05,Panter:07}, but are difficult to compare 
quantitatively as they differ in the selection details. The differences are
largely due to the very different survey volumes probed and methods of 
determining the star formation history. Juneau et al. (2005) reach lower SFRs 
determined from optical emission lines from sources in Gemini Deep Deep Survey 
(GDDS) over an area 20 times smaller than the work presented here, hence their
sample includes far fewer of the rarer, extremely high SFR objects that we 
find. \citet{Panter:07} use the data from the Sloan Digital Sky Survey Third 
Data Release (SDSS DR3 covering 
a far greater area than our survey, $\sim5200\,$deg$^2$) and model the 
galaxy spectra to determine the individual and global star formation 
histories. 

What Figure~\ref{fig:sfrdm} truly shows us is the fall of the contribution by 
massive galaxies to the global SFR density. We cannot rule out that the 
contribution of the least massive to the total SFRD remains approximately 
constant over this redshift range. As the most massive galaxies 
are capable of the highest SFRs they dominate the upper end of the luminosity 
function at the high redshift which we explore. It is likely that as these 
massive galaxies are capable of such high SFRs they use up their material to 
form stars very quickly and hence quash their star formation rate. 

\subsubsection{Selection effects}
\label{sec.se}

We are selecting our star forming galaxies on a SFR indicator that is 
completely unbiased with respect to obscuration and with little selection
on redshift beyond that implied by our radio flux density limit. At any 
redshift we are naturally detect just the most 
luminous, and hence highest SFR galaxies. One thing to consider is whether 
the apparent observed downsizing is simply an effect of missing low-mass 
galaxies in our high redshift bins. SFGs in our lowest mass bin only 
become un-detectable at $z\ge3$ in our $K-\,$band or IRAC data. Low mass
SFGs at these redshifts with SFRs high enough to be included in our 
sample would extremely high Specific SFRs ($>10^{-8}\,$yr$^{-1}$) and we argue 
in \S3.6.1 that we have are unlikely to have missed any such extreme sources.
Hence, the low detection rate of low mass galaxies in 
the high and intermediate redshift bins is not due to our detection 
limits. 

One might argue that the trend of fewer high stellar mass galaxies at lower
redshift may be a volume selection effect. However, we can compare the number
of SFGs selected from their high SFR to that expected purely on mass 
selection. In the lowest redshift bin for the $13^{\rm H}$ sample, 
$0.1\le z\le0.4$, we would expect to detect $\sim0.2$ galaxies in the highest 
mass bin in this volume 
simply using the fitted Schechter function to the local stellar mass 
function of SFGs from \citet{Panter:07} \citep[assuming no evolution from 
$z=0$,][]{Fontana:06}. This prediction is consistent with our result of no 
galaxies found in this bin. 
In the highest redshift bin we would expect to detect
$\sim40$ high mass galaxies using the same method as before, \ie\ assuming no 
evolution of the mass function. However, given that \citet{Fontana:06} suggest 
that this part of the mass function is not in place until $z\le1$ this 
estimate of $\sim40$ may be an upper-limit and hence is consistent with our 
result where we detect 31 high mass galaxies in this redshift bin. These rough 
estimates have to be treated as such as the whole of the highest mass bin is 
above the knee of the stellar mass function where the space density changes 
most rapidly and is hardest to measure accurately. However, it is encouraging 
to see that number of massive galaxies varies approximately as expected. The 
ratio of massive to low mass galaxies must, of course,  
decrease toward higher redshifts 
if there is hierarchical evolution in the mass function, but our selection on 
SFR shows the opposite trend with a higher ratio of high mass to low mass 
galaxies at higher redshift.

A further factor to consider is the choice of a single mass-to-light ratio 
for the entire SFG sample. Mass-to-light ratios can vary by over an order
of magnitude from active to quiescent galaxies, but most of our sample are 
clearly quite active by any standard, hence any variation in the ratio will be
much less than this amount. Photometric stellar masses used here are also
quite model dependent unlike direct dynamical measurements. There is quite
a debate concerning photometric stellar masses in the literature at the 
moment \citep[e.g.][]{Maraston:05}. 
We note, though, that relative photometric stellar masses are more accurate 
than absolute photometric stellar masses so the general trends examined in 
this section and the next are genuine cosmological effects.

\subsection{Characteristic Times}
\label{sec.ssfr}

\bfig
\psfig{file=ztc2.ps,width=8.4cm,angle=270}
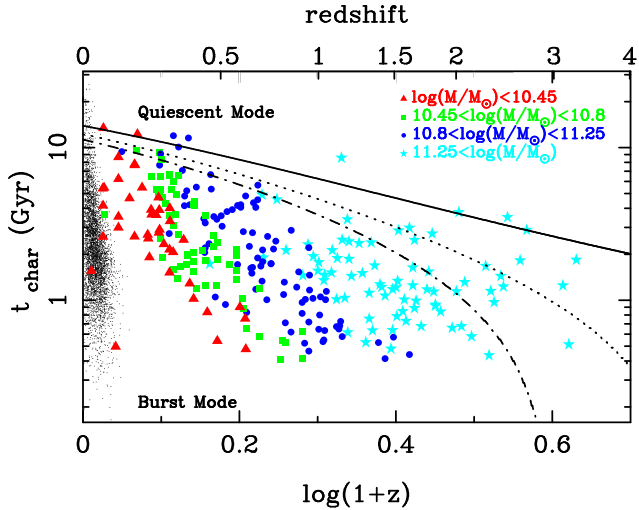
\caption{The ``characteristic time'' (the inverse of the SSFR) of the 
  radio star forming galaxies plotted against redshift. 
  The slowly decreasing solid line 
  indicates the age of the Universe at a given redshift and the dotted 
  (dot-dashed) line indicates the time since $z=5~(3)$. Galaxies below 
  these lines are forming stars at a faster rate than they were in the past 
  and are said to be in a ``burst mode'' and those galaxies above the line 
  are said to be in ``quiescent mode''. For reference, the Milky Way has a 
  characteristic time scale of $\sim100\,$Gyr putting it firmly in the 
  quiescent region of the figure (and off the plot), whilst M82 has a 
  characteristic time of $\sim0.7\,$Gyrs, well in the local burst region.}
\label{fig:ztc}
\efig

The specific star formation rate (SSFR) of a galaxy is defined as its 
star formation rate per unit stellar mass, and is representative of how active 
a galaxy is compared to its past properties. For all but the 
most massive galaxies our survey 
is biased against low SSFRs as we are selecting on SFR only. 
The inverse of the SSFR has been referred to as the {\it characteristic time 
scale} \citep[\eg\,][]{Juneau:05}. This number has units of time and gives an
idea of the nature of the star formation going on in the galaxy. For example 
if at a given redshift the characteristic time of a galaxy is less than the 
age of the Universe (or the time since the galaxy formed) then 
the current SFR is higher than the historic mean SFR and the galaxy can be 
said to be in a ``burst mode''. Conversely if the characteristic time is 
greater than the age of the Universe at a given time then the observed SFR 
is lower than the mean historic SFR and the galaxy can be said to be in a 
``quiescent'' mode. In Figure~\ref{fig:ztc} we plot the characteristic time 
as a function of redshift and along with the time since the Big Bang, $z=5$ 
and $z=3$. Most of our star forming
galaxies can be classified as being in ``burst mode'', although this 
classification becomes less certain at high redshift as it more strongly 
depends on the redshift of galaxy formation. This result implies that the current
observed star formation rates in most of the galaxies selected here are likely 
due to some trigger event beyond the initial gravitational collapse and 
formation of the galaxy, perhaps AGN activity or galaxy-galaxy interaction.
The fact that we detect the most SFGs in ``burst mode'' is 
simply a function of the depth of the survey of the parent sample. All but the 
lowest redshift sources have SFR $>10\,M_\odot$yr$^{-1}$, already an elevated 
value for most local galaxies. A much lower radio flux density limit is 
necessary to detect significant numbers of ``quiescent'' galaxies at any kind 
of distance.

\section{Conclusions}

We present the analysis of a multi-wavelength follow up to one of the
deepest radio surveys currently published. These data provide 
spectroscopic and photometric redshifts which are vital in discerning the 
nature of the sub-mJy radio population. With a philosophy of minimising 
the assumptions about the nature of this population we consider four
discriminators between AGN and SFGs based on purely the radio, or relative 
radio, properties of each source. These four discriminators are: radio 
morphology, radio spectral index, radio/near-IR and 
mid-IR/radio flux density ratios.
We find that 178 objects in the parent sample were classified as AGN by at 
least one of our AGN discriminators and $\sim90\%$ of these are classified 
by at least one of the flux ratio methods methods. 
Our radio-selected parent sample contains a roughly $40/60$ mix of AGN and 
SFGs, but with AGN dominating at high radio flux densities and SFGs dominating 
at faint flux densities. 
By recalculating the Euclidean normalised source counts 
by source type we find that the up-turn of the counts below 1\,mJy is mainly 
due to SFGs, but still with a $\sim25\%$ contribution from AGN at the 
faintest flux densities. This result is consistent with previous model fits 
to the source counts and estimates from ultra-deep MERLIN imaging.

We find many galaxies at high redshift which have SFRs exceeding those found 
in the local Universe and that these galaxies tended to be more massive than 
SFGs at lower redshift. 
We also find a trend of SFR correlating with 
stellar mass at all redshifts from the 6df-NVSS sample and the $13^{\rm H}$
sample, \ie\ for a galaxy at a given redshift and stellar mass there appears to
be an upper-limit to the SFR possible. 
After correcting for radio SFGs below our nominal detection limit 
($4\sigma=30\,\uJy$ at 1.4\,GHz), we derive the comoving star formation 
rate density as a function of redshift. We find our results are consistent
with those derived by methods at other wavelengths. 
We are able to look at the population of sources in the top end of the 
luminosity ($L_{\rm 1.4GHz}>10\times L_\star$) as a function of redshift 
and find that the typical mass of the sources making up the top one quarter of 
the star forming luminosity function changes dramatically with redshift. 
We also find that the {\it characteristic time} of the SFGs was generally
low implying an enhanced current SFR for most galaxies compared to their 
past mean SFR. 

The {\it Radio Morphology} and {\it Radio Spectral Index} discriminator 
methods should be just as powerful as the flux ratio methods 
if deeper high resolution and multi-wavelength data were available. The 
mid-IR/radio flux density ratio would be improved by deeper mid-IR data. 
Radio luminosity, not used as a discriminator in this work, will become a 
powerful discriminator when larger volumes are probed in deeper and wider 
surveys.
Unless the radio sources undetected at any other wavelength are extremely 
obscured (\ie\ $A_{\rm V}\ge 8$), then we are
able to reject the hypothesis that they are low-mass SFGs at $z\sim1$ because
none of these radio sources have detections in our sensitive $u$, $B$, and 
$g'-$band imaging. Such sources are most likely high redshift ($z>2$), obscured
type-2 AGN.

These results imply that deeper future radio surveys from eVLA,
eMERLIN, LOFAR and (eventually) SKA will mainly detect SFGs, not only high 
SFR objects at extreme redshifts, but also low luminosity, more quiescent SFGs 
at the redshifts probed in this work, giving a better view of the star 
formation in the distant Universe from radio data. The fraction of AGN is still
non-negligible hence discrimination between AGN and SFGs will be the principal 
challenge in exploiting future surveys.  This discrimination is by no means 
easy and, at least at present, largely statistical. Future surveys will need
to be backed up by multi-wavelength data,  particularly mid-IR and near-IR, 
photometry, to confidently separate the SFGs and AGN, as well as
deeper low frequency and high resolution radio data.

\section*{Acknowledgments}

We thank the referee for help improving the presentation of 
this paper. We thank V. Smol\v{c}i\'c, P. Capak, K. Sheth, M. Huynh and
R. Norris for useful discussions. We thank T. Mauch for help with the 
6dF-NVSS data set. We thank M. Jarvis for providing his model prior to 
publication. We thank K. Gunn much help in this project over the years.
AMH acknowledges support provided by the Australian Research Council
in the form of a QEII Fellowship (DP0557850).
This work is based in part on data obtained with the {\it Spitzer 
  Space Telescope}, which is operated by the Jet Propulsion Laboratory, 
California Institute of Technology under a contract with NASA. Support for 
this work was provided by an award issued by JPL/Caltech. Partial support 
for this work was provided by contract 1255094 issued by JPL/Caltech to 
the University of Arizona.

\bibliographystyle{mnras}

\appendix

\section{Redshift dependence of the radio-IR relation}

The radio-IR relation refers to the very close correlation, over five orders 
of magnitude, of the radio and IR luminosities of star forming galaxies 
\citep{Condon:91,Yun:02} as the radio and IR luminosities are directly 
related to star formation. To use radio data as a measure of star formation, 
we need to determine if the radio-IR relation, from which radio/SFR calibrations 
are ultimately derived, holds or changes above $z=1$ \citep[it has been shown 
to hold up to $z\sim1$ in the mid-IR,][]{Appleton:04}. In fact, two recent 
studies have indicated 
that a change is possible \citep{Kovacs:06,Vlahakis:07}. We proceeded 
somewhat differently from these studies to avoid two possible sources of bias. 
Firstly, we take a sample of radio-selected, high redshift galaxies 
(which have ULIRG or near-ULIRG luminosity) and a sample of local ULIRGs only, 
since the relation has not shown to be luminosity-independent into the 
ULIRG range (i.e. local samples from Yun et al. 2001 and Bell 2003 
are restricted to sub-ULIRG sources, $\log(L_{1.4GHz}/$WHz$^{-1})\le 24$).
\nocite{Yun:01, Bell:03} 
Secondly, we compared the radio and IR rest-frame wavelength fluxes between 
the local ULIRGs and high redshift galaxies, since free-free absorption 
can flatten the radio  spectrum and make extrapolations back to rest-frame 
1.4\,GHz uncertain. 

First, we assembled a local comparison sample of ULIRGs dominated by 
star formation \citep[according to the studies of][]{Farrah:03,
Armus:07} and with radio measurements at least at 1.4 and 8.4 
GHz \citep{Condon:91}. There are seven galaxies that meet these 
requirements; Arp 220, IRAS 1056, IRAS 1211, IRAS 1434, IRAS 1525, IRAS 
17208 and IRAS 2249. Figure~\ref{fig:sed} shows their spectral energy 
distributions, normalized at $260\,\um$. We over-plotted the measurements of 
\citet{Kovacs:06} of high redshift galaxies at 1.4\,GHz and 350 and $850\um$. 
We have not shown the measurements of galaxies at $z < 1.4$, and we 
also rejected measurements of three systems that are likely to be influenced 
by AGN and one more not detected at $350\,\um$: numbers 2, 3, 9, and 10 in 
their Table 1. Where they were available, we used the improved 1.4\,GHz
radio flux densities measurements from Biggs \& Ivison (2006) instead of those 
from Kovacs et al. (2006). 

\begin{figure}
  \caption{{\bf (This figure is available as a gif)}
    Observed rest-frame SEDs of local star forming ULIRGs (as indicated 
  in figure and normalised at $260\,\um$) compared to far-IR, submm and radio 
  observations of submm galaxies (SMGs, marked by dots) at 
  $1.4\le z\le5$ excluding SMGs with likely AGN. This plot indicates that 
  for a carefully selected sample of high redshift star forming ULIRGs
  the radio-IR luminosity relation show no obvious deviation from that 
  found locally.}
\label{fig:sed}
\end{figure}

We normalised the plotted points to provide a good fit at both 350 and 
$850\,\um$ (observed) to the SEDs of local ULIRGs. As Figure 1 shows, the 
radio and IR flux densities indicated for the high redshift galaxies 
($1.4\le z\le3.4$) agree well with the 
envelope of the SEDs of the local ULIRGs. In the radio, there is no apparent 
offset from the local galaxies, but the scatter is considerably larger than 
for the local ULIRGs. However, a significant part of this larger scatter is 
likely due to measurement errors, since many of the 1.4\,GHz flux densities 
have relatively low signal to noise. A further explanation for the scatter is
that the high redshift starbursts have a wider range of radio spectral 
indices. 

\bsp

\label{lastpage}

\end{document}